(Full file of the preprint)

\documentstyle[12pt]{article}
\begin{document}
\begin{center}
\vspace*{0.5cm}
{\Large\bf Spin-one ferromagnets with single-ion anisotropy in a
perpendicular external field}\\
\vspace*{0.7cm}
  Lei Zhou \\
\vspace*{0.2cm}
{\it Department of Physics, Fudan University,
Shanghai 200433, P. R.  China}\\
\vspace*{0.6cm}
Ruibao Tao\\
\vspace*{0.2cm}
{\it Center for Theoretical Physics, Chinese Center of Advanced
Science and
Technology (World Laboratory) ,
 P. O. Box 8730,
 Beijing 100080, China }\\
{\it and Department of Physics, Fudan University,
 Shanghai 200433, P. R. China}\\
\end{center}

\vspace*{1.0cm}

\noindent PACS numbers: 75.10Jm, 75.30Gw, 75.30Ds

\vspace*{1.0cm}

\centerline{\bf\large Abstract}

\vspace*{0.5cm}

In this paper, the conventional Holstein-Primakoff method is
generalized with the help of the characteristic angle transformation
[Lei Zhou and Ruibao Tao, J. Phys. A {\bf 27} 5599 (1994)]
for the spin-one magnetic systems with single-ion anisotropies.
We find that the weakness of the conventional method for such
systems can be overcome by the new approach. Two models will be
discussed to illuminate the main idea, which are the ``easy-plane"
and the ``easy-axis" spin-one ferromagnet, respectively. Comparisons
show that the current approach can give reasonable ground state
properties for the
magnetic system with ``easy-plane" anisotropy though the
conventional method never can, and can give a better representation
than the conventional one for the magnetic system
with ``easy-axis" anisotropy though the latter is usually believed to be
a good approximation in such case. Study of the easy-plane model
shows that there is
a phase transition induced by the external field, and the
low-temperature specific heat may have a peak as the
field reaches the critical value.

\newpage

\section{Introduction}

Magnetic systems with single-ion anisotropy $D(S^z_i)^2$ have been
attracting attentions for years since such kind of anisotropy was
found to be very popular in many magnetic materials
\cite{book1}-\cite{book2}.
On theoretical side, the spin-wave excitation in
such systems are not very easy to handle caused by the off-diagonal
effect of the single-ion anisotropy, especially when the spontaneous
magnetized direction is not the same as the anisotropic direction.
Many theoretical approaches have been developed to deal with such
kind of systems. Usually, one first apply a rotating transformation of
the spin vectors to determine the ground state, then perform a
Holstein-Primakoff (H-P) transformation to study the low-lying spin
wave excitations. However, this method was found to be a good
approximation only when the anisotropy is the ``easy-axis" case
(i.e. $D < 0$). In the ``easy-plane" case (i.e. $D > 0$), the method
was much worse. To understand it, one can study an easy-plane Heisenberg
model. If the conventional H-P method is used
naively to discuss the ground state and the magnon excitations
of such a system, an imaginary value of the excitation energy for
$``k=0"$ mode will always be encountered, which implies the
failure of this method.

Such deficiency of the conventional method is caused by missing an
important quantum effect. Actually, for the single-ion anisotropy
(no matter ``easy-axis" case or ``easy-plane" case), off-diagonal
term $D\sin^2\theta(S_i^{x'})^2$ will always appear in the Hamiltonian
as well as the diagonal terms $D\cos^2\theta(S_i^{z'})^2$
after introducing the spin vector rotation. Such off-diagonal term
may have the tendency to mix the single-site spin-state $|n\rangle$
with $|n+2\rangle$ and $|n-2\rangle$ to form the proper
eigenstates, and this spin-states mixing effect is completely
a quantum one which is very important in ``easy-plane" anisotropy
case. Unfortunately, such a quantum effect has been neglected
by the conventional H-P method. As the result, the conventional method
failed for the magnetic systems with ``easy-plane" anisotropy.

On the other hand, many methods have been proposed for the easy-plane
magnetic systems \cite{mme1}-\cite{zlca}. The matching of the
matrix elements (MME) method \cite{mme1}-\cite{mme2} was one which
can be used to consider the spin-states mixing effect
perturbatively so that it can give a reasonable
representation for an easy-plane ferromagnet when the single-ion
anisotropy is small \cite{mme1}-\cite{mme5},
and some numerical methods were developed for an
easy-plane spin-one ferromagnet \cite{nm91}-\cite{nm93}.
Recently, another method - the characteristic angle (CA)
method was proposed for the easy-plane spin-one ferromagnet
which could be applied to describe such spin-states mixing
effect by a variation parameter through a spin operator
transformation \cite{zlca}. The magnetic properties had been
investigated for such a system in zero field, and the results
seemed to be closer to the numerical results than those of the
MME method \cite{zlca}.

The present work is focused on generalizing the conventional H-P method
with the help of CA transformation for the spin-one magnetic systems
with single-ion anisotopies. Two particular models will be studied as the
illustration of the CA approach, although the latter is certainly not
limited to such models. The difficulties faced by the conventional H-P
method are overcome for such systems with the help of the new approach.

This paper is organized as follows. In the next section,
the easy-plane model is studied using the CA approach.
Detailed comparisons of the CA approach with the conventional method
is made in Section 3. Section 4 is devoted to an easy-axis model,
and the conclusions are summarized in the last section.

\section{Easy-plane case}

The first model we will study is an easy-plane spin-one ferromagnet
in an external magnetic field which is applied perpendicular to the
``easy-plane". The Hamiltonian of this system can be given as:
\begin{eqnarray}
 H=-J \sum_{ (i,j) } {\bf  S}_i \bullet {\bf  S}_j
        +D \sum_{i} ( S_{i}^{z} )^{2} - h\sum_i S^{z}_i,
\end{eqnarray}
where the first term is the exchange interaction, and the
second one is the single-ion anisotropy. The anisotropy
parameter $D$ is positive so that $x$-$y$ plane is
the so-called ``easy-plane" and $z$ axis is the ``hard axis".
An external magnetic field $h$ is applied along the ``hard axis".

Although the single-site part of Hamiltonian
$D( S_{i}^{z} )^{2} - h S^{z}_i$ has already been a
diagonalized form, it is still unreasonable to apply
a H-P transformation naively to discuss the magnetic
properties of such a system by assuming the ground state to
be the ordinary ferromagnetic state.
Actually if we do that, we will easily find that
the magnon excitation energy of $``k=0"$ mode will be always
negative in the case of $``h<D"$. That is because
the ``starting point" based on which the spin deviations
are discussed is wrong.

One must be very careful in finding a reasonable ``starting point".
Actually, in such a system, on the one hand, the spins are forced
into the ``easy-plane" by the single-ion anisotropy, on the other
hand, they have the tendency to point along the ``hard axis" caused
by the external field. As the result, this two effects must compete
with each other and a new direction $z'$ axis would be optimized to
describe the spontaneous magnetized direction. So, it's desirable to
introduce a new coordinates system (${\hat x'}$,${\hat y'}$,${\hat z'}$)
in which the spin components are related to those in the original
coordinates by the following transformation:
\begin{eqnarray}
S^{z}_i &=& \cos\theta_r S^{z'}_i - \sin\theta_r S^{x'}_i
\label{eq:lc1}\\
S^{x}_i &=& \cos\theta_r S^{x'}_i + \sin\theta_r S^{z'}_i\\
S^{y}_i &=& S^{y'}_i\label{eq:lc3}
\end{eqnarray}
Applying the above transformation to Hamiltonian (1), we have
\begin{eqnarray}
H = &-&J \sum_{ (i,j) } {\bf  S'}_i \bullet {\bf  S'}_j
       + D\cos^2\theta_r\sum_i (S^{z'}_i)^2
       + D\sin^2\theta_r\sum_i (S^{x'}_i)^2
       - h\cos\theta_r\sum_i S^{z'}_i\nonumber\\
&-& D\sin\theta_r\cos\theta_r\sum_i(S^{z'}_i S^{x'}_i
       + S^{x'}_i S^{z'}_i)
                + h\sin\theta_r\sum_i S^{x'}_i\label{eq:lch}
\end{eqnarray}
In a classical view, we can always determine $\theta_r$ based on
variation method assuming all spins are aligned along the $z'$
direction in the ground state. However, one should be careful in
quantum case, especially in the current ``easy-plane" anisotropy case.
Actually, if we apply the H-P transformation naively to investigate
the spin-waves excitation in such system, an imaginary value of the
magnon excitation energy for $``k=0"$ mode will always exist.
In fact, since
\begin{eqnarray}
D\sin^2\theta_r (S^{x'}_i)^2 = \frac{D}{4}\sin^2\theta_r (S_i^{+'}
S_i^{-'} + S^{-'}_iS^{+'}_i)
+ \frac{D}{4}\sin^2\theta_r (S^{+'}_iS^{+'}_i + S^{-'}_iS^{-'}_i),
\label{eq:off}
\end{eqnarray}
if the H-P transformation is applied naively to the Hamiltonian
(\ref{eq:lch}), one may find that the off-diagonal terms
$\frac{D}{4}\sin^2\theta_r (S^{+'}_iS^{+'}_i + S^{-'}_iS^{-'}_i)$
in the above equation have no contribution to the constant term
of the transformed Hamiltonian. That means the spin-states mixing
effect has already been neglected by the conventional H-P method.
Unfortunately, such effect is very important and must be considered
in such case. The characteristic angle (CA) transformation \cite{zlca}
 was developed
to describe the spin-state mixing effect in spin-one case
by introducing another variation parameter $\theta_c$:
\begin{eqnarray}
  S_{j}^{+'}&=&\cos\theta_c \tilde S_{j}^{+}
                 +\sin\theta_c  \tilde S_{j}^{-}
                 exp(i \pi \tilde S_{j}^{z}),\\
  S_{j}^{-'}&=&\cos\theta_c \tilde S_{j}^{-}
                 +\sin\theta_c exp(-i \pi \tilde S_{j}^{z} )
                 \tilde S_{j}^{+},\\
  S_{j}^{z'}&=&(1/2)[ S_{j}^{+'}, S_{j}^{-'} ]_{-}.
\end{eqnarray}
The spin operators are transformed to a new set of quasi-spin
operators $(\tilde S^{\pm}_j, \tilde S^z_j)$ which had been proved
to obey all spin-one operator's commutation rules \cite{zlca}.
After the CA transformation, we can apply a H-P transformation to
transform the quasi-spin operator to Bose one,
\begin{eqnarray}
 \tilde S_{i}^{z} &\longrightarrow& 1-a_{i}^{+}a_{i},\\
 \tilde S_{i}^{+} &\longrightarrow&
        \sqrt {2} \sqrt {1-(a_{i}^{+}a_{i}/2)} a_{i},\\
 \tilde S_{i}^{-} &\longrightarrow&
        \sqrt {2} a_{i}^{+} \sqrt {1-(a_{i}^{+}a_{i}/2)}.
\end{eqnarray}
Then, the Hamiltonian will have the following form:
\begin{eqnarray}
H = U_0 + H_1 + H_2 + \cdots\label{eq:hb}
\end{eqnarray}
where
\begin{eqnarray}
U_0 &=& N[-JZ\cos^22\theta_c + D - \frac{D}{2}\sin^2\theta_r
        (1+\sin2\theta_c) - h\cos\theta_r\cos2\theta_c]
        \label{eq:gs}\\
H_1 &=& - \frac{\sqrt{2}}{2} \sum_i [ D\sin\theta_r\cos\theta_r
        (\cos\theta_c + \sin\theta_c)
        - h\sin\theta_r(\cos\theta_c - \sin\theta_c) ]\nonumber\\
     & & ~~~~~~~~~~   (a^+_i + a_i)
\end{eqnarray}
and $H_2$ can be written in momentum ${\bf k}$ space as follows
\begin{eqnarray}
H_2 &=& \sum_k A_k a^+_k a_k
        + \sum_k B_k (a^+_ka^+_{-k} + a_ka_{-k})\\
A_k &=& 2JZ (\cos^2\theta_c - \gamma_k) + \frac{D}{2}\sin^2\theta_r
      (1+\sin2\theta_c)
        + h\cos\theta_r\cos2\theta_c\nonumber\\
    & &  ~~~~    - D\cos^2\theta_r,\label{eq:ak}\\
B_k &=& \frac{\sqrt{2}}{2} [-JZ\sin4\theta_c
        + \frac{D}{2}\sin^2\theta_r
        \cos2\theta_c - h\cos\theta_r\sin2\theta_c]\nonumber\\
    & &  ~~~~  + JZ\sin2\theta_c\gamma_k. \label{eq:bk}
\end{eqnarray}

Based on the variation method we understand that
the two parameters $\theta_r,\theta_c$ should be determined
by minimizing the ground state energy. As a first order
approximation, we may obtain:
\begin{eqnarray}
\frac{1}{N}\frac{d}{d\theta_r}U_0(\theta_r,\theta_c)
&=& -D\sin\theta_r\cos\theta_r(1+\sin2\theta_c) +
      h\sin\theta_r\cos2\theta_c = 0\label{eq:ro}\\
\frac{1}{N}\frac{d}{d\theta_c}U_0(\theta_r,\theta_c)
&=&4JZ\sin2\theta_c\cos2\theta_c - D\sin^2\theta_r\cos2\theta_c
    + 2h\cos\theta_r\sin2\theta_c\label{eq:ca}\nonumber\\
&=& 0
\end{eqnarray}
Eq. (\ref{eq:ro})
is just the same as the condition $H_1 = 0$ and Eq. (\ref{eq:ca}) can
cancel most of the off-diagonal terms which are in the
square bracket in the expression of $B_k$.
If we substitute the solution of the above non-linear
equations into to the Hamiltonian (\ref{eq:hb}),
then diagonalize the harmonic part of
Hamiltonian $H_2$ by a usual Bogolyubove transformation,
the total Hamiltonian will be:
\begin{eqnarray}
H= U'_0   + \sum_k E_k \alpha_k^+\alpha_k + \cdots
\end{eqnarray}
where
\begin{eqnarray}
U'_0 &=& U_0 - \frac{1}{2}\sum_k\ A_k +
\frac{1}{2}\sum_k\sqrt{A^2_k-4B^2_k}\label{eq:gse}\\
E_k &=& \sqrt{A^2_k - 4B^2_k}\label{eq:gap}.
\end{eqnarray}
The ground state in such a method can be defined by
\begin{eqnarray}
\alpha_k |0\rangle = 0
\end{eqnarray} 
Then the induced magnetization $M(h)$ is derived
in the harmonic approximation as follows:
\begin{eqnarray}
M(h)&=& \frac{1}{N}\sum_i\langle 0|S_i^{z}|0\rangle
    \simeq \frac{1}{N}\sum_i\cos\theta_r\langle 0|S_i^{z'}|
        0\rangle\nonumber\\
&\simeq& \frac{1}{N}\sum_i\cos\theta_r\{
      \cos2\theta_c \langle 0|\tilde S_i^z|0\rangle
      +\sin\theta_c\cos\theta_c
      \langle 0|(\tilde S_i^+)^2 + (\tilde S_i^-)^2|
        0\rangle\}\nonumber\\
&\simeq& \frac{1}{N}\sum_i\cos\theta_r\{
      \cos2\theta_c \langle 0|1-a^+_ia_i|0\rangle
      +\sqrt{2}\sin\theta_c\cos\theta_c
      \langle 0|a_i^{+2} + a_i^2|0\rangle\}\nonumber\\
&\simeq& \frac{3}{2}\cos\theta_r\cos2\theta_c
        -\frac{1}{2N}\sum_k\frac{\cos\theta_r\cos2\theta_c A_k
        + 2\sqrt{2}\cos\theta_r\sin2\theta_c B_k}
        {\sqrt{A^2_k - 4B^2_k}}.\label{eq:mag}
\end{eqnarray}

Thus, put the solution of Eqs. (\ref{eq:ro})-(\ref{eq:ca})
into Eqs. (\ref{eq:gse})-(\ref{eq:gap}), (\ref{eq:mag}),
such physical properties as the ground state energy, the magnon
dispersion relation and the induced magnetization can be obtained.
However, since it is very difficult to solve the non-linear
equations analytically, numerical calculations are carried out.
The system with anisotropy parameter $D/4JZ=0.6$ has been
studied as an example.

$\theta_r$ and $\theta_c$ as the functions of the external
field have been drawn together in figure 1 from which one can
find that they are both the decreasing function of the external
field. It is understood that $\theta_r$ is used to describe the
spontaneous magnetized direction and $\theta_c$
the spin-states mixing effect, in zero applied field case, the spontaneous
magnetized direction will be the $x$ axis ($\theta_r=90^{\circ}$) and
the spin-states mixing effect should be the strongest since the
off-diagonal term $D\sin^2\theta_r(S^{x'}_i)$ is the strongest,
and the value of $\theta_c$ is consistent with Ref. \cite{zlca}
where $h=0$ case has already been discussed.
While the external magnetic field is strengthened, the spins will
point along a direction which is closer to the $z$ axis due to
the interaction with the external field so that $\theta_r$ will
decrease, at the same time, the
spin-states mixing effect is also weakened since the off-diagonal
interactions in the total Hamiltonian will turn smaller along
with the decrement of $\theta_r$. However, when $h$ reaches a
critical value $h_c=D$, the external
magnetic field is so strong that the spins will not be rotated
any longer,
and the off-diagonal term comes to zero either, as the result,
$\theta_r$ and $\theta_c$ will vanish simultaneously.

\section{Comparisons and Discussions}

In this section, we will compare the CA method with the conventional
H-P method in details and discuss the magnetic properties of the above
mentioned system.

\vspace*{0.8cm}

First, one may find that more quantum effects have been comprised in
the constant term of the Hamiltonian by the new approach.

Introducing the H-P transformation naively to Hamiltonian (5),
the constant term can be found as:
\begin{eqnarray}
U_0^{HP} &=& N(-JZ + D\cos^2\theta_r - h\cos\theta_r)
+ N\frac{D}{2}\sin^2\theta_r \nonumber\\
&=& U_0^{C} + N\frac{D}{2}\sin^2\theta_r.\label{eq:lcgs}
\end{eqnarray}
where $U_0^C$ is the ground state energy obtained by a classical
rotating transformation.

After applying the CA transformation, the ground state energy is
Eq. (\ref{eq:gs}) which can be rewritten as:
\begin{eqnarray}
U_0 = U_0^{HP} + U_1
\end{eqnarray}
where
\begin{eqnarray}
U_1 &=& N(JZ\sin^22\theta_c - \frac{D}{2}\sin^2\theta_r\sin2\theta_c
+ 2h\cos\theta_r\sin^2\theta).
\end{eqnarray}
$U_1$ is an additional term introduced by the CA transformation
which will vanish as $\theta_c=0$.

The conventional H-P method is a semi-classical one.
The only quantum effect in (\ref{eq:lcgs}) is the term
$N\frac{D}{2}\sin^2\theta_r$ which comes from the contribution
of $\frac{D}{4}\sin^2\theta_r(S^{+'}_iS^{-'}_i + S^{-'}_iS^{+'}_i)$
in Eq. (\ref{eq:off}), and other terms which have been collected
in $U_0^{C}$ of Eq. (\ref{eq:lcgs}) can be easily recovered by a
classical method.
However, after applying the CA transformation, it can be clearly
found that there is an additional contribution $U_1$ in
the expression of $U_0$ which describes the single-site
spin-states mixing effect through
the variation parameter $\theta_c$.
Such an effect is completely a quantum
one which has no classical counterpart,
and it is expressed as the
competition of the exchange term ($JZ$)
and the external term $(h)$
with the single-ion anisotropy term ($D$).

So, more quantum effects have been considered
by the CA approach than the conventional H-P
method even in the constant term of the Hamiltonian.
Furthermore, considering such quantum effect will lead to a
lower ground state energy since the ground state
in CA method is selected to be the minimum point
of function $U_0$ although that in the conventional H-P method
is not so.

\vspace*{0.8cm}

Now, one can compare the CA method with the conventional H-P
method for the elementary excitation of such system.
Put the solution of $\theta_r$ and $\theta_c$ into
Eq. (\ref{eq:gap}), the magnon excitation gap can be calculated
with respect to the external magnetic field and the result has been
shown in figure 2, where one can find that the magnon excitation gap
obtained by the CA method will always be positive or zero.

However, based on the conventional H-P method,
one should obtain the variation parameter
$\theta_r$ by minimizing
(\ref{eq:lcgs}) which yields:
\begin{eqnarray}
\frac{1}{N}\frac{d}{d\theta_r}U_0^{HP}=
-D\sin\theta_r\cos\theta_r + h\sin\theta_r = 0
\end{eqnarray}
then substitute the solution of the above equation into
Eqs. (\ref{eq:ak}-\ref{eq:bk}), (\ref{eq:gap}), one can
easily find that the magnon excitation gap in H-P method will be
\begin{eqnarray}
\Delta^{HP} = \sqrt{\frac{h^2 - D^2}{2D}}.
\end{eqnarray}
Of cause, the excitation gap will never be real when $h < D$.
That is to say the conventional
H-P method can not be applied naively to study the magnetic
systems with ``easy-plane" anisotropy. However, the CA method
has overcome this difficulty as shown in figure 2.

\vspace*{0.8cm}

The induced magnetization as the function of the external
 magnetic field has been drawn in figure 3. From figures 1-3,
one may find that the point $h_c=D$ is very strange and there
seems to be a phase transition in such a point. As the external
field is strengthened across $h_c$, the system transits to a phase
in which the spins are not rotated any longer.
Actually, this phase transition can be clearly
shown by calculating the low-temperature specific heat $C_v$.

Suppose the system is at low temperature, and only the low energy
excitation is considered, then the inner energy will be:
\begin{eqnarray}
E(T)=E_0 + \displaystyle\sum_k E_k \frac{1}{exp(\frac{E_k}{K_BT})-1}
\end{eqnarray}
So the specific heat can be obtained:
\begin{eqnarray}
C_v = \frac{dE(T)}{dT} =(\frac{1}{N}
\displaystyle\sum_k\frac{(\frac{E_k}{K_B T})^2
exp({\frac{E_k}{K_B T}})}
{(exp({\frac{E_k}{K_B T}}) -1)^2}) NK_B
\end{eqnarray}
The specific heat of the system as the function of
the external magnetic field at the temperature $K_BT/JZ = 0.1$
has been shown in figure 4, in which a peak can be apparently
found in the critical point $h_c=D$. The physics can be understood as
follows: in the vicinity of the phase transition point $h_c$, the
magnons will be excited without a gap (fig. 2) so that the thermal
fluctuations are strong.

\section{Easy-axis case}

Now we will study another model which is an ``easy-axis"
spin-one ferromagnet in an external magnetic field whose direction is
perpendicular to the ``easy-axis". The Hamiltonian of such system
can be given as:
\begin{eqnarray}
 H=-J \sum_{ (i,j) } {\bf  S}_i \bullet {\bf  S}_j
        - D \sum_{i} ( S_{i}^{x} )^{2} - h\sum_i S^{z}_i,
\end{eqnarray}
where the single-ion anisotropy makes $x$ axis an easy-axis, and an
external magnetic field is applied along $z$ direction.

At a first glance, this Hamiltonian has a same classical picture
as the last model - the anisotropic interaction and the
interaction with external field have to compete with each other and
will be balanced at some angle $\theta_r$. So a rotating transformation
of the spin vectors is helpful. In fact, many authors have used
this method to discuss various kinds of magnetic systems with
single-ion anisotropy and believed this approximation will work
well when the single-ion anisotropy is ``easy-axis" case.
After the rotating transformation (\ref{eq:lc1})-(\ref{eq:lc3}),
the Hamiltonian will be:
\begin{eqnarray}
H = &-&J \sum_{ (i,j) } {\bf  S'}_i \bullet {\bf  S'}_j
       - D\sin^2\theta_r\sum_i (S^{z'}_i)^2
       - D\cos^2\theta_r\sum_i (S^{x'}_i)^2
       - h\cos\theta_r\sum_i S^{z'}_i\nonumber\\
&-& D\sin\theta_r\cos\theta_r\sum_i(S^{z'}_i S^{x'}_i
+ S^{x'}_i S^{z'}_i)
+ h\sin\theta_r\sum_i S^{x'}_i
\end{eqnarray}
there are still off-diagonal terms
$(-D\cos^2\theta_r\sum_i (S^{x'}_i)^2)$,
in the Hamiltonian, and this off-diagonal interactions
may be important in some cases. So it is
helpful to apply the CA transformation to get a
more reasonable representation.

Actually, after almost the same procedure as that for the
easy-plane model, the Hamiltonian can be transformed to:
\begin{eqnarray}
H = U_0 + H_1 + H_2 + \cdots
\end{eqnarray}
where
\begin{eqnarray}
U_0 &=& N[-JZ\cos^22\theta_c - D + \frac{D}{2}\cos^2\theta_r
        (1+\sin2\theta_c) - h\cos\theta_r\cos2\theta_c]\\
H_1 &=& - \frac{\sqrt{2}}{2} \sum_i [ -D\sin\theta_r\cos\theta_r
        (\cos\theta_c + \sin\theta_c)
        + h\sin\theta_r(\cos\theta_c - \sin\theta_c) ]\nonumber\\
     & & ~~~~~~~~~~   (a^+_i + a_i)
\end{eqnarray}
and
\begin{eqnarray}
H_2 &=& \sum_k A_k a^+_k a_k
        + \sum_k B_k (a^+_ka^+_{-k} + a_ka_{-k}),\\
A_k &=& 2JZ (\cos^22\theta_c - \gamma_k) - \frac{D}{2}\cos^2\theta_r
      (1+\sin2\theta_c)
        + h\cos\theta_r\cos2\theta_c\nonumber\\
    & &    ~~~~  + D\sin^2\theta_r,\label{eq:esak}\\
B_k &=& \frac{\sqrt{2}}{2} [-JZ\sin4\theta_c
        - \frac{D}{2}\cos^2\theta_r
        \cos2\theta_c - h\cos\theta_r\sin2\theta_c]\nonumber\\
    & &    ~~~~+ JZ\sin2\theta_c\gamma_k.\label{eq:esbk}
\end{eqnarray}
The two variational parameters $\theta_r,\theta_c$ satisfy:
\begin{eqnarray}
\frac{1}{N}\frac{d}{d\theta_r}U_0(\theta_r,\theta_c)
&=& -D\sin\theta_r\cos\theta_r(1+\sin2\theta_c) +
h\sin\theta_r\cos2\theta_c = 0\label{eq:es}\\
\frac{1}{N}\frac{d}{d\theta_c}U_0(\theta_r,\theta_c)
&=&4JZ\sin2\theta_c\cos2\theta_c + D\cos^2\theta_r\cos2\theta_c
+ 2h\cos\theta_r\sin2\theta_c\nonumber\\
&=& 0
\end{eqnarray}
where Eq. (\ref{eq:es}) will cancel
the $H_1$ part of the Hamiltonian.

Physical properties such as magnon excitation, the
induced magnetization have the same forms as which in last model
(i.e. Eqs. (\ref{eq:gse}-\ref{eq:gap}),(\ref{eq:mag}))
except the concrete expression of the functions $A_k$ and $B_k$.

Very similar to the ``easy-plane" case, the constant term in the
Hamiltonian can further be divided into two two terms:
\begin{eqnarray}
U_0 = U_0^{HP} + U_1
\end{eqnarray}
where
\begin{eqnarray}
U_0^{HP} &=& N(-JZ - D\sin^2\theta_r - h\cos\theta_r)
- N\frac{D}{2}\cos^2\theta_r \nonumber\\
&=& U_0^{C} - N\frac{D}{2}\cos^2\theta_r\label{eq:eslc}\\
U_1 &=& N(JZ\sin^22\theta_c + \frac{D}{2}\cos^2\theta_r\sin2\theta_c
+ 2h\cos\theta_r\sin^2\theta_c).
\end{eqnarray}
$U_0^{HP}$ is the contribution of the conventional H-P method, and $U_1$ is
an additional term which describes the quantum effect of spin-states
mixing in single-site. All the discussions are similar to the first
model: more quantum effects have been comprised
in the constant term of the Hamiltonian by the CA method,
and considering such effect in CA method will lead to a lower
ground state energy than that in the conventional H-P method. Actually,
as shown in figure 5 where $U_0$ and $U_0^{HP}$ are drawn together
with respect to the external magnetic field for an easy-axis spin-one
ferromagnet, $U_0$ is always found to be lower than $U_0^{HP}$.

Now it is interesting to compare the elementary excitations
calculated by the CA method with those by the conventional H-P method.
In the latter case, $\theta_C=0$ and $\theta_r$ is obtained
by minimizing $U_0^{HP}$ (\ref{eq:eslc}) which yields:
\begin{eqnarray}
\frac{1}{N}\frac{d}{d\theta_r}U_0^{HP}=
-D\sin\theta_r\cos\theta_r + h\sin\theta_r = 0
\end{eqnarray}
So, substitute the solution of the above equation
back into Eqs. (\ref{eq:esak}-\ref{eq:esbk}) then
into Eq. (\ref{eq:gap}), the elementary excitation in the
conventional H-P method can be calculated readily.

The magnon excitation gaps have been calculated by
both methods with respect to the external magnetic field,
and the results are presented in figure 6 where the solid
line is by CA method and the dotted line is by the conventional
H-P method. From the figure one may find that: when the
external field is close to the anisotropy parameter $D$,
there is a small region where the magnon excitation
gap calculated by the H-P method will be imaginary, which
indicates that this approximation is poor in such
area. However, the solid line in the figure tells us that CA
method has overcome this difficulty and the magnon excitation gap
will always be real and positive in CA method. Actually,
as shown in figure 7 where the values of the
two variation parameters $\theta_c$ $\theta_r$ are
drawn together with respect to the external field,
when $h$ is close to $D$, $\theta_r$ comes to zero and
$\theta_c$ becomes somewhat larger indicating that the
spin-state mixing effect caused by the off-diagonal interaction
may be very strong. So, we must consider
such effect with the help of CA transformation in such
case, otherwise, the ``starting point"  may be unreasonable
and will lead to an imaginary minimum excitation energy. Outside
this region, the off-diagonal terms are not so strong comparing
to the diagonal parts, as the result, the spin-states mixing effect
is not very drastic and the conventional H-P method might be a
reasonable approximation as many authors believed. However,
the CA transformation may always be helpful to
get a more reasonable representation for such a system.

\section{Conclusions}

To summarize, in this paper, the conventional method has been
generalized with the help of the characteristic angle transformation for
the spin-one magnetic systems. The difficulties faced
by the conventional H-P method for magnetic systems with single-ion
anisotropy has been overcome by the new approach. Two models have been
discussed to illuminate the main ideas, of which one is an easy-plane
spin-one ferromagnet in an external field applied perpendicular to the
``easy-plane", and the other is an easy-axis spin-one ferromagnet in
an external field applied perpendicular to the ``easy-axis".
Comparisons between the new approach and the old one
show: more quantum effects have been considered by the CA
method, as the result, CA method can examine the ground state
properties of the ``easy-plane" spin-one ferromagnet
although the old method never can, and CA method can give an improved
representation for the ``easy-axis" spin-one ferromagnet although the
conventional H-P method is usually believed to be a good approximation
in such case. Also, study of the easy-plane model shows that a
phase transition may take place induced
by the applied field, and the low-temperature specific heat
is found to have a peak when the external field reaches the
critical value.

\vspace{1cm}

\noindent {\bf Acknowledgments}

\vspace*{0.3cm}

\noindent This research is supported by National Science Foundation
of China and the National Education commission under the grant for
training doctors.

\newpage

\vspace*{0.5cm}

\noindent{\bf\large Captions:}

\vspace{0.5cm}

\noindent Figure 1: $\theta_r$, $\theta_c$ as the functions of
the external field
for the easy-plane spin-one ferromagnet with $D/4JZ=0.6$.

\vspace{0.5cm}

\noindent Figure 2: Magnon excitation gap as the function of the
external field
for the easy-plane spin-one ferromagnet with $D/4JZ=0.6$.

\vspace{0.5cm}

\noindent Figure 3: Induced magnetization as the function of the
external field
for the easy-plane spin-one ferromagnet with $D/4JZ=0.6$.

\vspace{0.5cm}

\noindent Figure 4: Specific heat as the function of the external
 field at the
temperature $K_BT/JZ=0.1$
for the easy-plane spin-one ferromagnet with $D/4JZ=0.6$.

\vspace{0.5cm}

\noindent Figure 5: Ground state energy with respect to the external
magnetic field in the case of using the conventional H-P method
$U_0^{HP}$ and using the CA method for an easy-axis spin-one
ferromagnet with anisotropic parameter $D/4JZ=0.3$.

\vspace{0.5cm}

\noindent Figure 6: Magnon excitation gaps of the easy-axis spin-one
ferromagnet with $D/4JZ=0.3$
as the functions of the external magnetic field for the case using the
CA method (solid line) and using the conventional H-P method
(dotted line).

\vspace{0.5cm}

\noindent Figure 7: $\theta_r$, $\theta_c$ as the functions of the
external field
for the easy-axis spin-one ferromagnet with $D/4JZ=0.3$.


\newpage

\begin{thebibliography}{99}
\vspace*{1.0cm}
\bibitem{book1}  {\it Crystalline Electric Field and Structure Effect in
                f-Electron Systems},edited by Crow J E,
                Gruertin R P, and  Mihalisin T W (Plenum, New York, 1980)
\bibitem{book2}   {\it Crystalline Electric Field Effects in f-Electron
                Magnetism}, edited by Guertine R P, Suski W, Zolnierek
                Z (Plenum, New York, 1982)
\bibitem{mme1}    P. A. Lindg\.{a}rd and A. Kowalska, J. Phys. C: Solid St.
              Phys. {\bf 9} 2081 (1976)
\bibitem{mme2}    P. A. Lindg\.{a}rd and O. Danielsen, J. Phys. C: Solid St.
              Phys. {\bf 7} 1523 (1974)
\bibitem{mme3}    J. F. Cooke and P. A. Lindg\.{a}rd, Phys. Rev. {\bf B16} 408
              1978 J. Appl. Phys. {\bf 49} 2136 (1977)
\bibitem{mme4}     U. Balucani, M. G. Pini, A. Rettori and V. Tognetti
               J. Phys. C: Solid St. Phys. {\bf 13} 3895 (1980)
\bibitem{mme5}    E. Rastelli and P. A. Lindg\.{a}rd, J. Phys. C: Solid St.
              Phys. {\bf 12} 1899 (1979)
\bibitem{nm91}    C. F. Lo, K. K. Pan and Y. L. Wang, J. Appl. Phys.
                {\bf 70}(10) 6080 (1991)
\bibitem{nm93}    K. K. Pan  and Y. L. Wang, J. Appl. Phys.
              {\bf 73} (10) 1099 (1993).
\bibitem{zlca}    Lei Zhou and Ruibao Tao,
               J. Phys. A: Math and Gen. {\bf 27} 5599 (1994).
\end{thebibliography}
\end{document}